\documentclass[conference]{IEEEtran}
\IEEEoverridecommandlockouts
\usepackage{amsmath,amssymb,amsfonts}
\usepackage{listings}
\usepackage{algorithm}
\usepackage{algorithmicx}
\usepackage[noend]{algpseudocode}
\usepackage{graphicx}
\usepackage{textcomp}
\usepackage{xcolor}
\usepackage{tikz}
\usetikzlibrary{shapes,positioning,calc,arrows,chains}
\usepackage[unicode, hidelinks]{hyperref}
\usepackage[hyphenbreaks]{breakurl}
\usepackage[style=ieee, bibencoding=utf8, citestyle=numeric-comp]{biblatex}
\def\BibTeX{{\rm B\kern-.05em{\sc i\kern-.025em b}\kern-.08em
    T\kern-.1667em\lower.7ex\hbox{E}\kern-.125emX}}
\usepackage{booktabs}
\usepackage{colortbl}
\usepackage{multirow}
\usepackage{flushend}
\usepackage{verbatim}
\usepackage[inline]{enumitem}

\usepackage{subcaption}
\usepackage{mathtools}

\algnewcommand{\algorithmicand}{\textbf{ and }}
\algnewcommand{\algorithmicor}{\textbf{ or }}
\algnewcommand{\OR}{\algorithmicor}
\algnewcommand{\AND}{\algorithmicand}

\definecolor{light-gray}{gray}{0.80}

\begin{document}

\title{Crash Report Accumulation During Continuous Fuzzing}

\author{
\IEEEauthorblockN{
  Ilya Yegorov\IEEEauthorrefmark{1}\IEEEauthorrefmark{2} and
  Georgy Savidov\IEEEauthorrefmark{1}\IEEEauthorrefmark{2}
}
\IEEEauthorblockA{
  \IEEEauthorrefmark{1}Ivannikov Institute for System Programming of the RAS
}
\IEEEauthorblockA{
  \IEEEauthorrefmark{2}Lomonosov Moscow State University
}
Moscow, Russia \\
\{Yegorov\_Ilya, avgor46\}@ispras.ru
}

\maketitle

\begin{tikzpicture}[remember picture, overlay]
\node at ($(current page.south) + (0,0.65in)$) {
\begin{minipage}{\textwidth} \footnotesize
 \copyright~2024 IEEE. Personal use of this material is permitted. Permission
 from IEEE must be obtained for all other uses, in any current or future media,
 including reprinting/republishing this material for advertising or promotional
 purposes, creating new collective works, for resale or redistribution to
 servers or lists, or reuse of any copyrighted component of this work in other
 works.
\end{minipage}
};
\end{tikzpicture}

\begin{abstract}
Crash report accumulation is a necessary step during continuous fuzzing. Dynamic
  software analysis techniques like fuzzing and dynamic symbolic execution
  generate a large number of crashes for analysis. However, the time and
  resource constraints often lead to the postponement of fixing some less
  critical issues, potentially introducing new errors in future releases. Thus,
  there is a need to distinguish new errors from old ones. We propose a crash
  accumulation method and implemented it as part of the CASR~\cite{casr}
  toolset. We evaluated our approach on crash reports collected from fuzzing
  results.
\end{abstract}


\section{Introduction}

Nowadays the number of lines of code in modern projects is growing
rapidly~\cite{codebase}. At the same time, the number of existing software
errors also increases. Security development life cycle helps to reduce the
amount of potential vulnerabilities contained in the release version of a
product. Various dynamic software analysis techniques, such as coverage-guided
continuous fuzzing or dynamic symbolic execution, can produce a huge number of
crashes. Their subsequent processing may require automation tools for crash
analysis, such as CASR~\cite{casr}, Igor~\cite{igorek} or
Cluster-fuzz~\cite{clusterfuzz}, which allow to evaluate the similarity of
crashes and classify them into clusters that represent unique errors.

After examination of obtained results, developers make changes to the software
product aimed at fixing existing problems. However, the correction of some
of them (perhaps not the most critical ones), due to the lack time and resources,
may be postponed until the next versions of the program, which may also include
some new errors. And this cycle can be repeated a considerable number of times.
Once again, when receiving a set of crashes, it would be good to know
which of them belong to those that analyzed previously, and which will be
considered new, introduced by the latest updates. Thus, it is necessary to
update the clustering structure so that some of crashes would be merged
into old clusters, while others would form new ones.

The article is organized as follows. Section II describes approaches related to
handling crashes. Section III provides an overview of various methods for
accumulating crashes during the development process. Section IV presents
a description of our chosen method and experimental results. Section V
covers some implementation details. Section VI concludes the article.

\section{Related Work}\label{sec:rw}

\subsection{Casr-Cluster}\label{sec:casr-cluster}

\subsubsection{Common Analysis Approach}

\begin{figure}[h]
  \center{\includegraphics[width=0.5\linewidth]{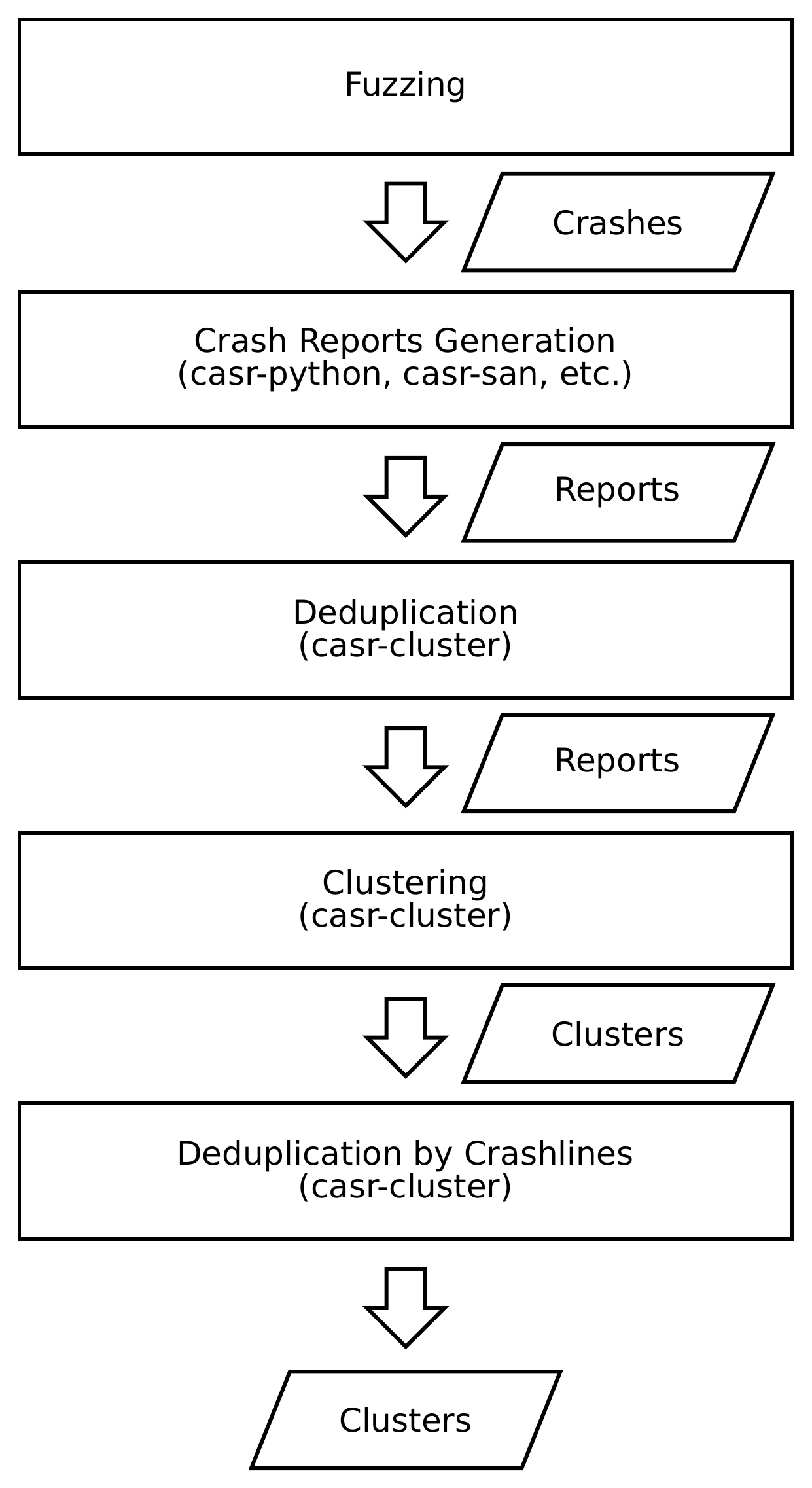}}
  \caption{Standard crash analysis scenario scheme}
  \label{fig:casr-pipeline}
\end{figure}

CASR~\cite{casr} is a open-source toolchain for crashes and exceptions analysis.
Casr-Cluster is a CASR console application for deduplication and clustering
crashes. Analyzing crashes with Casr-Cluster usually involves several
consecutive steps (fig. \ref{fig:casr-pipeline}):
\begin{itemize}
  \item Crash report generation,
  \item Report deduplication,
  \item Report clustering,
  \item Report deduplication inside clusters by crashline.
\end{itemize}

The Casr-Cluster analysis algorithm is based on the analysis of stacktraces in
program code~\cite{casr-cluster, vishnyakov22-sydr-fuzz}.

\subsubsection{Clustering}\label{sec:clustering}

The following value is used as the distance between two clusters:
\begin{align}
  dist(a, b) = 1 - similarity(a, b)
\end{align}
Where $similarity(a, b)$ is custom Casr-Cluster pseudometric.

Hierarchical clustering~\cite{hierarchy} is started based on the stacktrace
distance matrix. The distance between two clusters is defined as the maximum of
the pairwise distance between crashes retrieved from the two clusters:
\begin{align}\label{dist}
  dist(CL_i, CL_j) = \underset{a \in CL_i, b \in CL_j}{max(dist(a, b))}
\end{align}

Where $CL_i$ and $CL_j$ are different clusters.

\begin{figure}[h]
  \center{\includegraphics[width=0.5\textwidth]{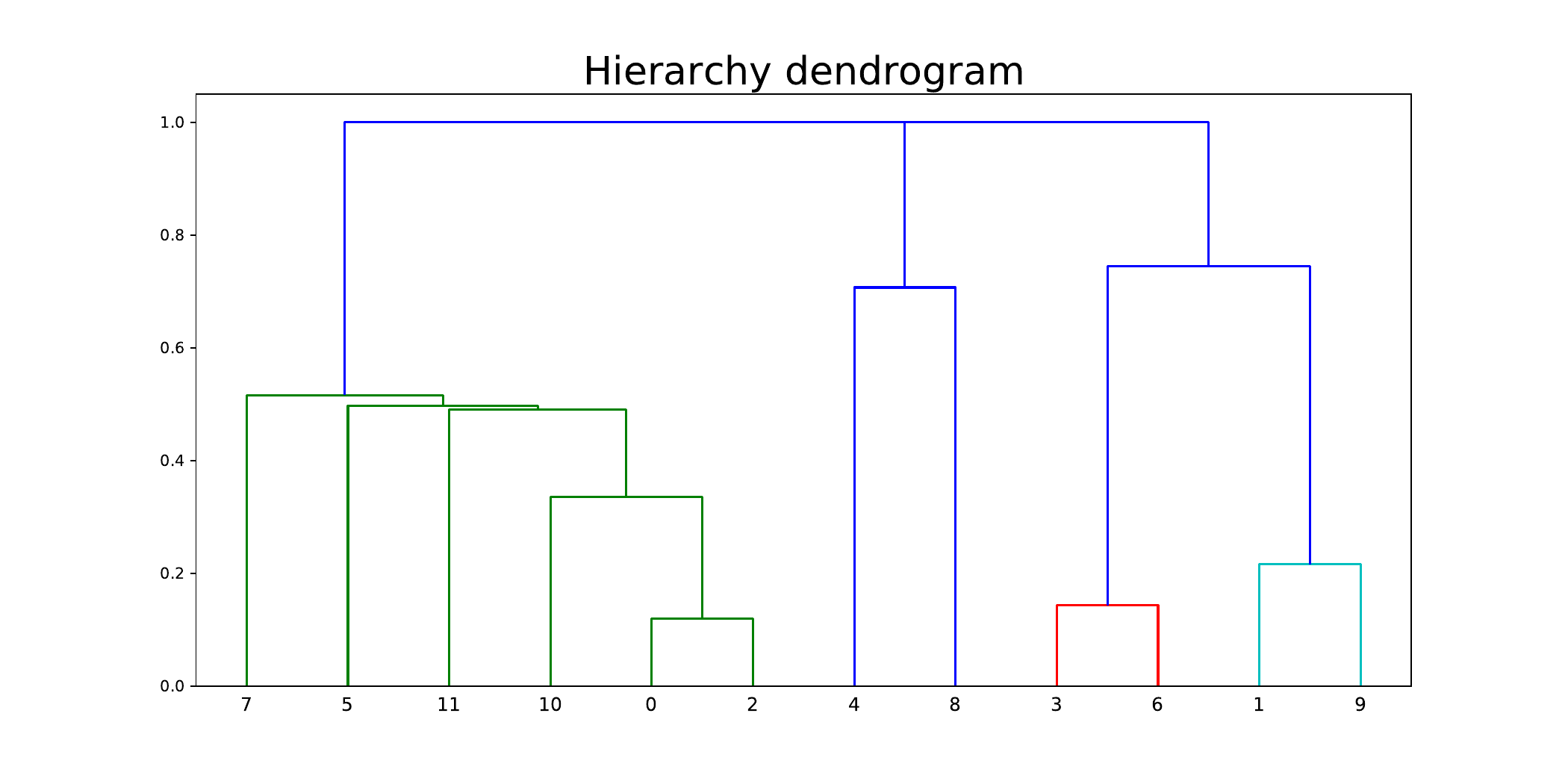}}
  \caption{Hierarchical clustering dendrogram}
  \label{fig:dendro}
\end{figure}

After obtaining hierarchical clustering dendrogram (fig. \ref{fig:dendro}) flat
clusters are created by merging subtrees of dendrogram until specified threshold
(\textit{THRESHOLD}) of distance between clusters (\ref{dist}) is reached. It's
clearly that \textit{THRESHOLD} imposes cluster diameter restrictions.

\subsubsection{Deduplication}

The \textit{similarity} coefficient is used for deduplication. Two crashes are
considered the same if $similarity = 1$ (that is, all frames in both call stacks
are the same).

\subsubsection{Deduplication by Crashline}

In case if several crashes in same cluster have same crashlines these crashes is
considered the same.

\subsection{ClusterFuzz}

ClusterFuzz~\cite{clusterfuzz} is a scalable fuzzing infrastructure. Google uses
ClusterFuzz to fuzz all of its products and as a back-end fuzzing feature in
OSS-Fuzz~\cite{oss-fuzz}. ClusterFuzz provides many features that help integrate
fuzzing into the software project development process:
\begin{itemize}
  \item Highly scalable: can run on cluster of any size.
  \item Accurate deduplication of crashes.
  \item Supports multiple coverage guided fuzzing engines
    (libFuzzer~\cite{libfuzzer}, AFL~\cite{afl}, AFL++~\cite{fioraldi20} and
    Honggfuzz~\cite{hongg}) for optimal results (with ensemble fuzzing and
    fuzzing strategies).
\end{itemize}

First, the stacktraces (more precisely, their upper parts) are checked for
complete matches. If stacktraces are differ, the largest common frame
subsequence is searched between them. Two crashes are considered identical if
the length of such subsequence exceeds the \textit{SAME\_FRAME\_THRESHOLD}
coefficient. Otherwise, for $\forall i \in range(min(|stacktrace_1|,
|stacktrace_2|))$
\begin{align}
  similarity\_ratio_i = \frac{|f_{1,i}| + |f_{2,i}| - d(f_{1,i}, f_{2,i})}{|f_{1,i}| + |f_{2,i}|},
\end{align}
is considered, where $f_{1,i}$ is the i-th frame in the first call stack
($stacktrace_1$), $f_{2,i}$ is the i-th frame in the second call stack
($stacktrace_2$), $|stacktrace_i|$ is the number of frames in $stacktrace_i$ ($i
= \overline{1,2}$), $|f_{j,i}|$ is the length of $f_{j,i}$ as a string ($j =
\overline{1,2}$), and $d( S_1, S_2)$ is Levenshtein distance~\cite{leven}. After
that, the arithmetic mean of all $similarity\_ratio_i$ is calculated, and if it
overcomes a certain coefficient (\textit{COMPARE\_THRESHOLD}), crashes are again
considered the same, otherwise different.

\subsection{Igor}

Igor~\cite{igorek} is a crash deduplication tool, that uses program execution
trace to calculate similarity between two crashes. Igor generates a Control-Flow
Graphs to represent the execution flow of each program, and then applies the
Weisfeiler-Lehman Subtree Kernel algorithm to distinguish test cases with
different root causes. This algorithm creates a similarity matrix, that is used
in Spectral Clustering method. Igor does not use call stack to calculate the
similiraty.

\section{Possible Approaches}

First of all let's introduce some concept notation. \\
Let's identify crashes with their stacktraces (or just traces) in accordance
with the fact that the analysis is based on them. \\
Let $\mathbb{TR}$ be a set of all possible traces, $\mathbb{CL = CL(TR)}$ be a
set of all possible clusters composed of traces from $\mathbb{TR}$ and $Clusters
\subseteq \mathbb{CL}$ be a given set of old clusters.

\subsection{Reclustering Approach}\label{sec:reclustering}

\begin{figure}[h]
  \centering
  \begin{subfigure}{.5\textwidth}
    \centering
    \fbox{\includegraphics[height=.12\textheight]{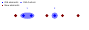}}
    \caption{Input}
    \label{fig:recluster1}
  \end{subfigure}
  \begin{subfigure}{.5\textwidth}
    \centering
    \fbox{\includegraphics[height=.12\textheight]{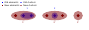}}
    \caption{Result}
    \label{fig:recluster2}
  \end{subfigure}
  \caption{Reclustering and mapping example}
\end{figure}

The first idea we considered is just to recluster existing crashes according to
(\ref{sec:clustering}) approach. The drawback of this method is that it could
destroy old cluster structure. However, we can recluster all new and old crashes
and try to map new clusters onto the old ones. This means that we are looking
for old clusters that are subclusters of new ones. Example could be found in
fig. \ref{fig:recluster1}. As result of reclustering and mapping (fig.
\ref{fig:recluster2}) we can see that cluster $1$ is subset of cluster $1`$ and
cluster 2 is subset of cluster 2`, so it seems logical to add new crashes from
1` to 1, from 2 to 2` and save crash from 3` as new cluster.

\begin{figure}[h]
  \centering
  \begin{subfigure}{.5\textwidth}
    \centering
    \fbox{\includegraphics[height=.12\textheight]{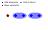}}
    \caption{Input}
  \end{subfigure}
  \begin{subfigure}{.5\textwidth}
    \centering
    \fbox{\includegraphics[height=.12\textheight]{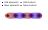}}
    \caption{Result}
  \end{subfigure}
  \caption{Bad reclustering example}
  \label{fig:bad-recluster}
\end{figure}

Unfortunately, clustering algorithm is sensitive to input data, i.e. resulting
clusters can differ significantly. For example, adding only one element totally
changes entire cluster structure as it's presented at fig.
\ref{fig:bad-recluster}.

So this approach cannot be used in our case.

\subsection{Minimal Diameter Approach}\label{sec:mindiam}

The second idea being considered is picking the cluster for updating with
particular crash based on their "closeness". The one way is to choose cluster
with minimal resulting diameter. Of course, we need to ensure that resulting
diameter is lower than \textit{THRESHOLD}, because we want the updated and
created clusters to be similar to the originally generated
(\ref{sec:clustering}).
 
So we get following approach: \\
For each $trace \in \mathbb{TR}: \exists Cluster \in Clusters:$
\begin{align}
  diam (Cluster \cup \{trace\}) < THRESHOLD
\end{align}
we choose
\begin{align}
  Cluster = \underset{Cluster\in Clusters}{Argmin}diam(Cluster\cup\{trace\})
\end{align}
for updating. All the remaining ones are clustered separately.

\begin{figure}[h]
  \centering
  \begin{subfigure}{.5\textwidth}
    \centering
    \fbox{\includegraphics[height=.12\textheight]{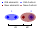}}
    \caption{Choosing minimal diameter}
    \label{fig:mindiam1}
  \end{subfigure}
  \begin{subfigure}{.5\textwidth}
    \centering
    \fbox{\includegraphics[height=.12\textheight]{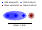}}
    \caption{Leaving trace separately due to the large diameter}
    \label{fig:mindiam2}
  \end{subfigure}
  \caption{Minimum diameter example}
\end{figure}

For example, at fig. \ref{fig:mindiam1} new element will be added to cluster
$2$, because $d1 > d2$, and at fig. \ref{fig:mindiam2} new one will be saved as
new cluster, because possible resulting diameter when adding to existing cluster
will be greater than \textit{THRESHOLD}.

\begin{figure}[h]
  \centering
  \fbox{\includegraphics[height=.12\textheight]{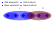}}
  \caption{Bad minimum diameter example}
  \label{fig:bad-mindiam}
\end{figure}

But there are cases when cluster updating will not be good enough. For example,
we can look at fig. \ref{fig:bad-mindiam}. Resulting diameter when adding new
element to cluster $1$ will be greater than it when adding to cluster $2$, so it
will be added to $2$, but it's full duplicate of some element from $1$. Of
course, it is not good behavior for accumulating algorithm. It may seem that
taking duplicate is slightly exaggerated example, but if we take element that
slightly differs from it we will get the same situation with cluster
"diffusion". Thus, this approach also cannot be used.

\section{Suggested Approach}\label{sec:sugg-appr}

The last approach (\ref{sec:mindiam}) showed that first of all it would be a
good idea to check weather the specified trace is "inside" some existing
cluster. So we have come up the idea to divide traces into several groups before
analysis.

\subsection{Trace Grouping Approach}\label{sec:loyal}

\subsubsection{Traces Groups}

Let $trace \in \mathbb{TR}$ and $Cluster \in \mathbb{CL}$ and let
\begin{itemize}
  \item $trace \in Dup(Cluster) \xLeftrightarrow{def}$ \\
    $\exists trace' \in Cluster: dist(trace, trace') = 0$
  \item $trace \in Inner(Cluster) \xLeftrightarrow{def}$ \\
    $trace \notin Dup(Cluster) \wedge diam(Cluster \cup \{trace\}) =
    diam(Cluster)$
  \item $trace \in Outer(Cluster) \xLeftrightarrow{def}$ \\
    $diam(Cluster) < diam(Cluster \cup \{trace\}) < THRESHOLD$
  \item $trace \in Oot(Cluster) \xLeftrightarrow{def}$ \\
    $diam(Cluster \cup \{trace\}) \geq THRESHOLD$
\end{itemize}

It's easy to see that for $\forall trace \in \mathbb{TR}$ and $\forall Cluster
\in \mathbb{CL}$ $trace$ will belong to one and only one set of $Dup(Cluster)$,
$Inner(Cluster)$, $Outer(Cluster)$, $Oot(Cluster)$ sets. Thus, $\mathbb{TR}$ is
decomposed into direct sum of related sets:
\begin{align}\label{trace-sets}
\scriptsize
\begin{split}
  \mathbb{TR} &= Dup(Cluster) \\
  & \oplus Inner(Cluster) \\
  & \oplus Outer(Cluster) \\
  & \oplus Oot(Cluster)
\end{split}
\end{align}

This means that each trace belongs to one and only to one of the groups relative
to given cluster.

\begin{figure}[h]
  \centering
  \fbox{\includegraphics[height=.20\textheight]{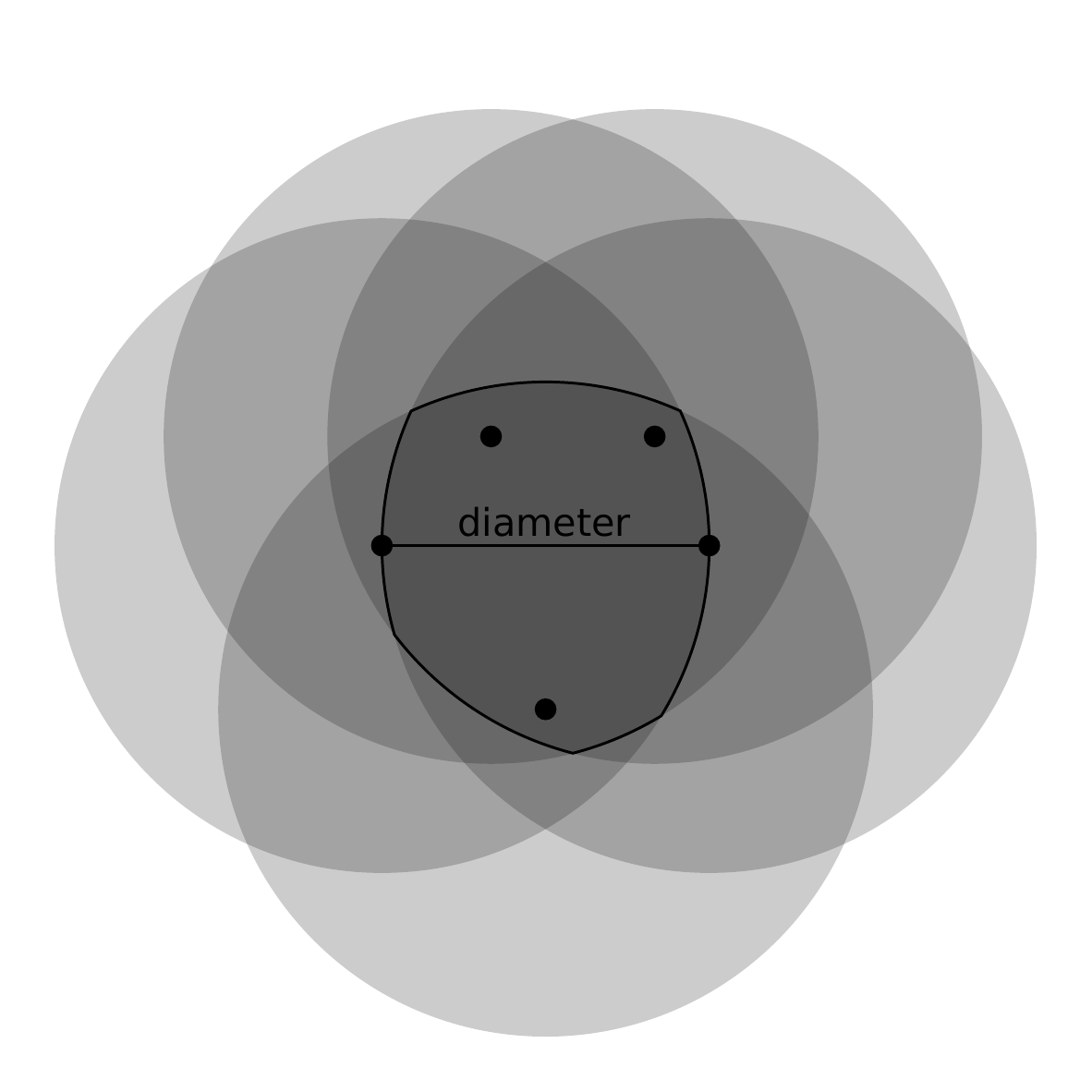}}
  \caption{Visualization of $Inner$ set}
  \label{fig:inner}
\end{figure}

In essence, this definition of $Inner$ set specifies the set of all traces lying
"inside" the cluster, i.e. potentially belonging to it. $Inner$ set may be
represented as intersection of spheres, where each sphere has a cluster element
as its center and a radius equal to the diameter of the cluster, without
existing traces (points of existing traces specifies $Dup$ set) (fig.
\ref{fig:inner}).

\subsubsection{Dealing with Groups}\label{sec:main}

Before we get into the approach itself, let's define several support sets:
\begin{itemize}
  \item $Dup = \{trace \in \mathbb{TR}|$ \\
    $\exists Cluster \in Clusters: trace \in Dup(Cluster)\}$
  \item $Inners(trace) = \{Cluster \in Clusters|$ \\
    $trace \in Inner(Cluster)\}$
  \item $Outers(trace) = \{Cluster \in Clusters|$ \\
    $trace \in Outer(Cluster)\}$
  \item $OOT = \{trace \in \mathbb{TR}|$ \\
    $\forall Cluster \in Clusters: trace \in Oot(Cluster)\}$
\end{itemize}
According to (\ref{trace-sets}), it's obvious that for $\forall trace \in
\mathbb{TR}$
\begin{align}
\scriptsize
\begin{split}
  & \biggl(trace \in Dup \vee Inners(trace) \neq \emptyset \vee Outers(trace) \neq
  \emptyset \biggr) \: \oplus \\
  & \oplus \: trace \in OOT
\end{split}
\end{align}
Thus, globally we can distinguish two cases:
\begin{itemize}
  \item $trace \in OOT$
  \item $trace \notin OOT$ or, what is equivalent,
\end{itemize}
\begin{align}\label{no-oot}
\scriptsize
\begin{split}
  trace \in Dup \vee Inners(trace) \neq \emptyset \vee Outers(trace) \neq \emptyset
\end{split}
\end{align}

In the first case we have set of traces quite far from all other clusters
according to $OOT$ definition. So we can simply cluster separately all traces
from $OOT$ to new clusters the same way as it was suggested in \ref{sec:mindiam}
for remaining traces.

In the second case we need to go through all options according to (\ref{no-oot})
with following priority for $newtrace$:
\begin{itemize}
  \item If $newtrace \in Dup \Rightarrow$ just shed it out
  \item Else if $Inners(newtrace) \neq \emptyset \Rightarrow$ \\
    add new trace to some $Cluster \in Inners(newtrace)$
  \item Else if $Outers(newtrace) \neq \emptyset \Rightarrow$ \\
    add new trace to some $Cluster \in Outers(newtrace)$
\end{itemize}

After that we should cluster all traces from $OOT$ and add result as new
clusters.

Thereby, we get approach that has no disadvantages of \ref{sec:mindiam}
approach.

\subsubsection{Cluster Collision Resolution}

\begin{figure}[h]
  \centering
  \fbox{\includegraphics[height=.16\textheight]{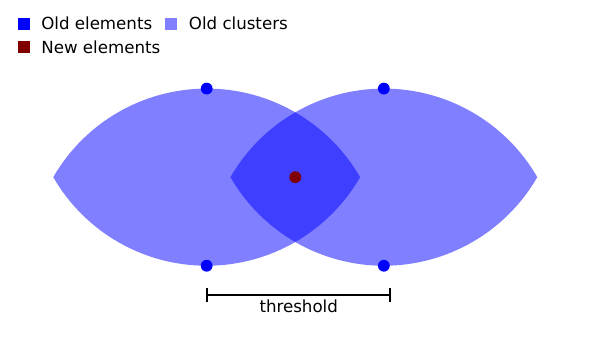}}
  \caption{Inners collision example}
  \label{fig:collision}
\end{figure}

Suggested approach (\ref{sec:main}) would be work only if it's clear what
cluster from $Inners(newtrace)$ or $Outers(newtrace)$ we need to pick. It's
trivial when the sets contain only one element. But if they don't? What we need
to do if $|Inners(newtrace)| > 1$ or $|Outers(newtrace)| > 1$? An example of
such case can be seen at fig. \ref{fig:collision}.

Suggested solution is supposed to be some sort of conservative. This means that
clusters should grow at a minimum speed. Taking it into account, we considered
several ways to choose cluster for adding:
\begin{enumerate}
  \item Diam: $\underset{Cluster \in Clusters}{Argmin}diam(Cluster \cup \{trace\})$
  \item Delta: $\underset{Cluster \in Clusters}{Argmin}$ \\
    $|diam(Cluster \cup \{trace\}) - diam(Cluster)|$
  \item Dist: $\underset{Cluster \in Clusters}{Argmin}dist(Cluster, \{trace\})$
\end{enumerate}

Note, than second strategy is not suitable for multi inners case, because value
in brackets is always equal to zero in that case.

\begin{figure}[h]
  \centering
  \begin{subfigure}{.5\textwidth}
    \centering
    \fbox{\includegraphics[height=.12\textheight]{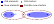}}
    \caption{Diam-Delta/Diam-Dist counterexample}
    \label{fig:counterex1}
  \end{subfigure}
  \begin{subfigure}{.5\textwidth}
    \centering
    \fbox{\includegraphics[height=.12\textheight]{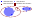}}
    \caption{Delta-Dist counterexample}
    \label{fig:counterex2}
  \end{subfigure}
  \caption{Equality counterexamples}
\end{figure}

The question then arises, maybe we can choose more general from them or maybe
some of them is equal to another or its subcase? Unfortunately, the answer to
all these questions is no. At fig. \ref{fig:counterex1} we can see, that (here
and further values with $`$ mean that they belong to new possible clusters)
\begin{align}
\begin{split}
  diam_1 = 1, \: diam'_1 = 3, \: \Delta_1 = 2, \: dist_1 = 2 \\
  diam_2 = 3, \: diam'_2 = 4, \: \Delta_2 = 1, \: dist_2 = 1
\end{split}
\end{align}
And thus,
\begin{align}
\begin{split}
  & diam'_1 < diam'_2 \wedge \Delta_1 > \Delta_2 \Rightarrow \\
  & Diam \nRightarrow Delta \wedge Diam \nLeftarrow Delta
\end{split} \\
\begin{split}
  & diam'_1 < diam'_2 \wedge dist_1 > dist_2 \Rightarrow \\
  & Diam \nRightarrow Dist \wedge Diam \nLeftarrow Dist
\end{split}
\end{align}
Same for fig. \ref{fig:counterex2}:
\begin{align}
\begin{split}
  & diam_1 = \sqrt{2} \approx 1.4, \: diam'_1 = 2, \\
 & \Delta_1 = 2 - \sqrt{2} \approx 0.6, \: dist_1 = \sqrt{2} \approx 1.4 \\
  & diam_2 = 1, diam'_2 = 2, \Delta_2 = 1, dist_2 = 1
\end{split} \\
\begin{split}
  & \Delta_1 < \Delta_2 \wedge dist_1 > dist_2 \Rightarrow \\
  & Delta \nRightarrow Dist \wedge Delta \nLeftarrow Dist
\end{split}
\end{align}

Based on these results, it was decided to choose the best strategy by experiment
evaluation.

\subsubsection{Evaluation and Intermediate Results}

For evaluating cluster quality we used \textit{Silhouette score}~\cite{sil},
since this problem involves the use of clustering without labels and the number
of clusters is not fixed. We divided crash corpus into two parts in different
ways and averaged the results, after that we compared our approach with
reference solution, i.e. with reclustering of full corpus. The results can be
seen in tab. \ref{tab:loyal_res}.
\begin{table*}[h]
  \centering
  \includegraphics[width=\textwidth]{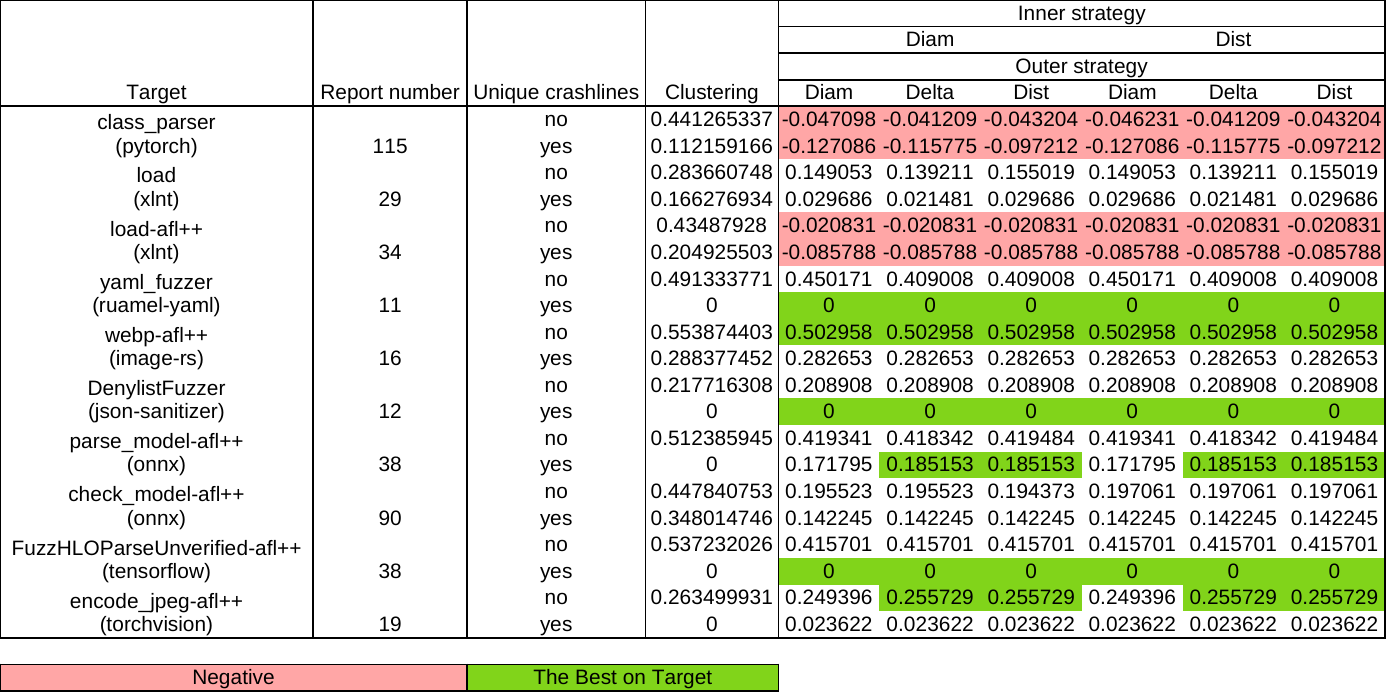}
  \caption{Suggested approach silhouette score}
  \label{tab:loyal_res}
\end{table*}
The table also highlights cells with a negative score and cells with the best
score for each target.

For several fuzz targets with zero score for reclustering we can see that
suggested solution has a better scores. However, it is important to note that
from an expert point of view, the clusters turned out to be very plausible. Also
we can see that choosing \textit{Inner strategy} does not influence the results.
This can mean that trace rarely falls into $Inner$ compared to other sets.

\subsection{Strong Conditions Approach}\label{sec:hard}

In tab. \ref{tab:loyal_res} we can see a lot negative scores. This section is
dedicated to solving this problem.

\subsubsection{Dealing with Negative Results}

\begin{figure}[h]
  \centering
  \begin{subfigure}{.25\textwidth}
    \centering
    \fbox{\includegraphics[height=.16\textheight]{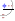}}
    \caption{Input}
    \label{fig:negscore}
  \end{subfigure}%
  \begin{subfigure}{.25\textwidth}
    \centering
    \fbox{\includegraphics[height=.16\textheight]{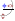}}
    \caption{Proper clustering}
    \label{fig:hard}
  \end{subfigure}
  \caption{Negative score example}
\end{figure}

To understanding the reasons of negative scores let's look at fig.
\ref{fig:negscore}: All elements inside \textit{THRESHOLD} area will be added to
old cluster, i.e. points $1$ and $2$ will be in different clusters, but it would
be more proper to put them in same cluster as it's presented at fig.
\ref{fig:hard}. In other words, point $1$ is more suitable for cluster with
point $2$, but silhouette score exactly indicates that an object is similar to
other objects in the cluster and not similar to objects from other clusters.

Thus, to avoid negative scores we can update existing clusters only with
elements from theirs $Inner$ sets, and deal with $Outer$ traces the same way as
with $OOT$ traces, i.e. separately cluster them all.

\subsubsection{Tolerance Level}

Having different approaches to dealing with $Outer$ traces we got the idea of
\textit{Tolerance level}: it's "degree" of how tolerant are old clusters to
differing new traces, i.e. $Outer$ traces. For the initially proposed approach
\ref{sec:main} we could say that it has \textit{Loyal} tolerance level, because
clusters with this level may be updated with any trace meeting the
\textit{THRESHOLD} condition. The new one \ref{sec:hard} will have \textit{Hard}
tolerance level, because only $Inner$ traces may be added to it.

\subsubsection{Intermediate Results}

In tab. \ref{tab:hard_res} we can see results of \ref{sec:hard} approach.
Notice that there is no \textit{Outer strategy} row, because we do not perform
analysis for $Outer$ traces and just cluster them with $OOT$ traces. Absolute
values of score in the table have become smaller than similar values in tab.
\ref{tab:loyal_res}. The reason why it happened is that resulting clusters turn
out to be very small with \textit{Hard level}. Of course, there are fewer
negative scores and the absolute values of them have become smaller, but
absolute values of positive scores have become smaller too.

\begin{table}[t]
  \centering
  \includegraphics[width=.5\textwidth]{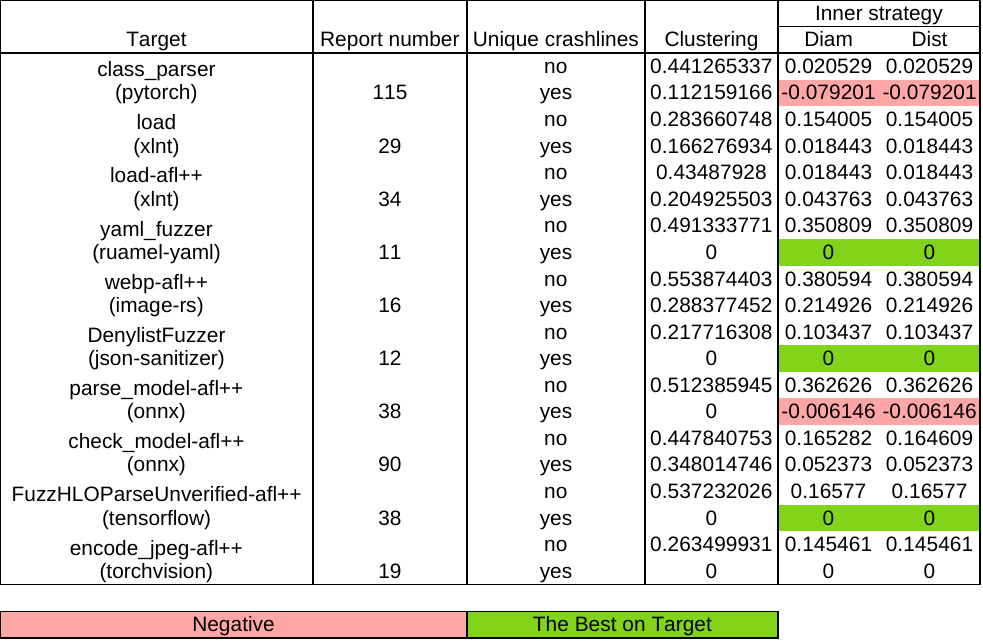}
  \caption{Hard level approach silhouette score}
  \label{tab:hard_res}
\end{table}

\subsection{Combined Approach}\label{sec:soft}
\begin{figure}[H]
  \centering
  \fbox{\includegraphics[height=.16\textheight]{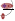}}
  \caption{Soft level clustering}
  \label{fig:soft}
\end{figure}

Cluster structure at fig. \ref{fig:hard} seems pretty good, but may be enhanced
like at fig. \ref{fig:soft}: excluding cluster with points $1$ and $2$ we merged
all others into single one, because such cluster fits within \textit{THRESHOLD}.

Thus, we can improve our approach: perform \textit{Hard level} accumulation
firstly, and after that try to merge new clusters with old ones. For the new
approach we will say that it has \textit{Soft} tolerance level.

\begin{table}[t]
  \centering
  \includegraphics[width=.5\textwidth]{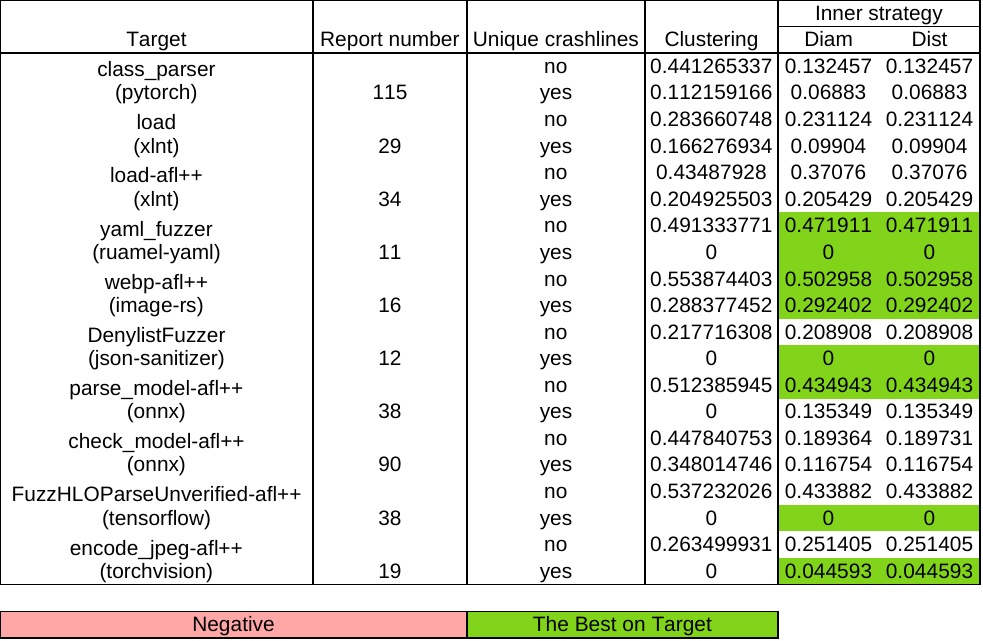}
  \caption{Soft level approach silhouette score}
  \label{tab:soft_res}
\end{table}

The results of such approach can be seen in tab. \ref{tab:soft_res}. First of
all, we can see that there is no negative scores at all. Secondly, it can be
noticed that with \textit{Soft level} we have the best result for almost all
fuzz targets for now. This is not surprising, since this approach is a
combination of the best and most successful ideas of all other approaches.

\subsection{Hierarchical Approach}\label{sec:hier}
In both \ref{sec:hard} and \ref{sec:soft} approaches we clustered $Outer$ traces
separately disregarding information about old clusters and their elements, but
it might be useful to it. So we get the idea to cluster old clusters with
$Outer$ traces together, but unlike \ref{sec:reclustering} approach we will use
old clusters as single, indivisible elements. Also due to the preliminary
$Inner$ traces dealing we won't have cases like at fig. \ref{fig:bad-mindiam}
during clustering. After clustering we have two cases for an $Outer$ trace:
\begin{enumerate*}
  \item cluster containing the trace also contains some old cluster;
  \item cluster containing the trace contains only new traces.
\end{enumerate*} \\
In first case we just need to add the trace to that old cluster. In the second
case we save that cluster as new one.

It should also be noted that it is impossible for two old clusters to end up
together in a new one if they were obtained as valid result of
\ref{sec:clustering} algorithm.

The tab. \ref{tab:hier_res} shows the results of such approach. It's clear that
approach scores are very close to original cluster algorithm results: everywhere
except two rows the difference is less than 0.1; and also the scores are the
best among the rest for almost all fuzz targets.

\begin{table}[h]
  \centering
  \includegraphics[width=.5\textwidth]{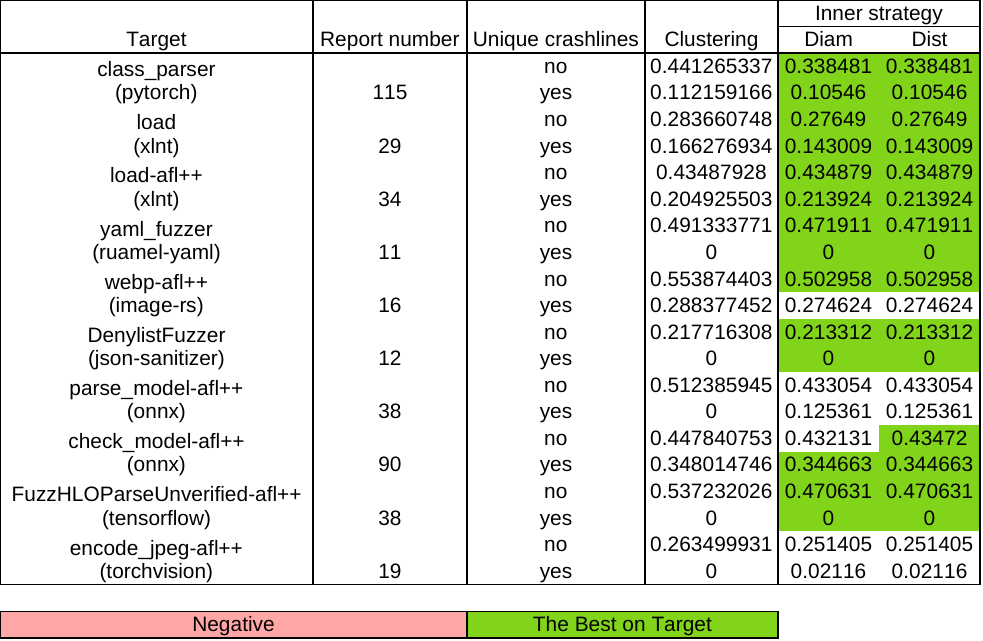}
  \caption{Hierarchy approach silhouette score}
  \label{tab:hier_res}
\end{table}

\section{Implementation}

Suggested method (\ref{sec:sugg-appr}) has been implemented as functionality of
the Casr-Cluster~\cite{casr} tool. As for approaches, hierarchical clustering
(\ref{sec:hier}) and \textit{Diam} inner strategy were chosen due to the results
(tab.  \ref{tab:loyal_res}, \ref{tab:hard_res}, \ref{tab:soft_res},
\ref{tab:hier_res}) and slight differences between \textit{Diam} and
\textit{Dist} strategies and ease of implementation of \textit{Diam} to compared
with \textit{Dist}.

\section{Conclusion}
We propose and implement crash accumulation method, based on hierarchical
clustering. The method could be applied to crash reports collected via Casr tool
on Linux operating systems. The method allows to determine which of the newly
received crashes were analyzed previously, and does not destroy the old cluster
structure, which simplifies further work on crash analysis.

\printbibliography

@inproceedings{fioraldi20,
    title={{{AFL++}}: Combining Incremental Steps of Fuzzing Research},
    author={Fioraldi, Andrea and Maier, Dominik and Ei{\ss}feldt, Heiko and Heuse, Marc},
    booktitle={14th USENIX Workshop on Offensive Technologies (WOOT 20)},
    year={2020},
    url={https://www.usenix.org/system/files/woot20-paper-fioraldi.pdf},
    language = {english},
}

@inproceedings{casr-cluster,
    title = {{{Casr-Cluster}}: Crash Clustering for Linux Applications},
    author = {Savidov, Georgy and Fedotov, Andrey},
    booktitle = {2021 Ivannikov ISPRAS Open Conference (ISPRAS)},
    pages = {47--51},
    year = {2021},
    organization = {IEEE},
    doi = {10.1109/ISPRAS53967.2021.00012},
}

@inproceedings{leven,
    author={Levenshtein, Vladimir Iosifovich},
    booktitle={Doklady Akademii Nauk SSSR},
    title={Dvoichnye kody s ispravleniem vypadenii i vstavok simvola},
    year={1965},
    volume={163},
    pages={845–848},
    language = {russian},
    url = {https://www.mathnet.ru/php/archive.phtml?wshow=paper&jrnid=dan&paperid=31411&option_lang=eng}
}

@inproceedings{vishnyakov22-sydr-fuzz,
    title = {{{Sydr-Fuzz}}: Continuous Hybrid Fuzzing and Dynamic Analysis for
        Security Development Lifecycle},
    author = {Vishnyakov, Alexey and Kuts, Daniil and Logunova, Vlada and
        Parygina, Darya and Kobrin, Eli and Savidov, Georgy and Fedotov,
        Andrey},
    booktitle = {2022 Ivannikov ISPRAS Open Conference (ISPRAS)},
    pages = {111--123},
    year = {2022},
    publisher = {IEEE},
    doi = {10.1109/ISPRAS57371.2022.10076861},
}

@conference {oss-fuzz,
    author = {Kostya Serebryany},
    title = {{OSS-Fuzz} - Google{\textquoteright}s continuous fuzzing service
        for open source software},
    year = {2017},
    address = {Vancouver, BC},
    publisher = {USENIX Association}
}

@article{sil,
title = {Silhouettes: A graphical aid to the interpretation and validation of cluster analysis},
journal = {Journal of Computational and Applied Mathematics},
volume = {20},
pages = {53-65},
year = {1987},
issn = {0377-0427},
doi = {https://doi.org/10.1016/0377-0427(87)90125-7},
url = {https://www.sciencedirect.com/science/article/pii/0377042787901257},
author = {Peter J. Rousseeuw}
}

@article{igorek,
  title={Igor: Crash Deduplication Through Root-Cause Clustering},
  author={Jiang, Zhiyuan and Jiang, Xiyue and Hazimeh, Ahmad and Tang, Chaojing and Zhang, Chao and Payer, Mathias},
  year={2021}
}

@misc{afl,
    title = {{AFL}: American Fuzzy Lope},
    url = {https://github.com/google/AFL}
}

@misc{hongg,
    title = {{Honggfuzz}: A security oriented, feedback-driven, evolutionary,
        easy-to-use fuzzer},
    url = {https://github.com/google/honggfuzz}
}

@misc{libfuzzer,
    title = {{{libFuzzer}}: A Library For Coverage-Guided Fuzz Testing},
    url = {https://llvm.org/docs/LibFuzzer.html}
}

@misc{hierarchy,
    title = {Hierarchical clustering},
    url = {https://en.wikipedia.org/wiki/Hierarchical_clustering}
}

@misc{casr,
    title = {{{Casr}}: Collect crash reports, triage, and estimate severity},
    url = {https://github.com/ispras/casr}
}

@misc{clusterfuzz,
    title = {{{ClusterFuzz}} is a scalable fuzzing infrastructure that finds
        security and stability issues in software.},
    url = {https://google.github.io/clusterfuzz}
}

@misc{codebase,
  title = {Codebases: Millions of lines of code},
  url = {https://www.informationisbeautiful.net/visualizations/million-lines-of-code}
}
\end{document}